\documentstyle[12pt,draft]{article}
\newcommand \be {\begin{equation}}
\newcommand \bea {\begin{eqnarray} \nonumber }
\newcommand \ee {\end{equation}}
\newcommand \eea {\end{eqnarray}}
 
 \newcommand \s {\sigma}

\begin{document}   

\title{Dynamical Behaviour of Low Autocorrelation Models}

\author{Gabriele Migliorini$^{(a)}$ and 
Felix Ritort$^{(a,b)}$\\
{\small $(a)$: Dipartimento di Fisica}\\
{\small Universit\`a di Roma {\it Tor Vergata},}\\
{\small Viale della Ricerca Scientifica, 00133 Roma, Italy}\\
{\small $(b)$: Dipartimento di Fisica and INFN,}\\
{\small Piazzale Aldo Moro, 2}\\
{\small 00187 Roma, Italy}\\[0.5em]}
\date {July 1994}
\maketitle

\begin{abstract}
We have investigated the nature of the dynamical behaviour in low
autocorrelation binary sequences. These models do have a glass
transition $T_G$ of a purely dynamical nature. Above the glass
transition the dynamics is not fully ergodic and relaxation times
diverge like a power law $\tau\sim (T-T_G)^{-\gamma}$ with $\gamma$
close to $2$.  Approaching the glass transition the relaxation
slows down in agreement with the first order nature of the
dynamical transition.  Below the glass transition the system exhibits
aging phenomena like in disordered spin glasses. We propose the aging
phenomena as a precise method to determine the glass transition and its
first order nature.
\end{abstract}

\vfill
\begin{flushright}
{\bf Preprint ROM2F/94/29}\\
{\bf cond-mat/9407105}\\
\end{flushright}
\vfill
\newpage

\section {Introduction} 

There has been much recent interest in the field of condensed matter
physics in the study of frustrated models
without explicit disorder \cite{I,II,BoMe,FrHe,CuKuPaRi}. These models
(also called deterministic models) show a very similar behaviour to spin
glasses \cite{LIBRO1}, i.e. there exist a very large number of
metastable configurations where the system remains trapped and it is
very difficult to reach the global equilibrium state in a dynamical
process starting from a random initial configuration.

The main difference between these frustrated models and spin glasses is
that in case of deterministic models the quenched disorder is not
present.  Because some symmetries are preserved in the deterministic
model, it is possible in some cases to explicitely construct the ground
state.  This possibility is generally forbidden in disordered systems
because no symmetry is preserved. Also, in disordered systems, each
realization of the randomness yields a different ground state implying
that there is much difficulty in devising any kind of algorithm to
identify the ground state.

Recently it has been shown that the application of techniques initially
devised for this random systems promises to be a powerful tool in the
understanding of the deterministic models \cite{I}. In particular, much
effort has been recently devoted to the study of the Bernasconi model
\cite{BERNAS}. This is an optimization problem in which one searches
for strings of binary digits with minimal autocorrelation. The 
high-temperature phase of this model has been exactly solved in the
particular case of periodic boundary conditions (\cite{I} hereafter
referred as paper I).  The system shows a static transition to a frozen
phase where the entropy is nearly zero. In the original Bernasconi model
with open boundary conditions an exact solution for the high $T$ phase
is still lacking but some approximations suggest that a similar static
transition takes also place in that case. This static transition is
different from the dynamical transition one observes in a real system.
The dynamical transition is higher than the static transition and
corresponds to the situation in which the system remains trapped in
metastable configurations. Below this dynamical transition temperature,
thermal fluctuations are very small and reminds a lot of what happens in
real glasses \cite{glass_book}. In the context of models without
explicit disorder this transition has been called the ${\em
glass~transition}$.  Starting a dynamical evolution from the
high-temperature region an enough large system is unable to see the
static transition because it gets trapped in the metastable phase at a
higher temperature.  For all purposes, it is always this higher
temperature transition which governs the dynamics. Within the realm of
disordered systems this dynamic transition can be computed using the
{\it marginality condition}
\cite{SOKODO}. This condition corresponds to the search for certain
saddle points of the free energy (not true maxima like in the static
case) such that one particular eigenvalue of the stability matrix
vanishes (the so called replicon eigenvalue). This condition corresponds
to the temperature at which dynamical stability disappears.  The
dynamical transition temperature has been obtained by several autors
using the dynamic mean-field theory of spin glasses initially studied by
Sompolinsky and Zippelius for the study of the SK model \cite{SOMZIP}.
Always in the framework of disordered systems, studies of Kirkpatrick
and collaborators on the $p$-spin Ising models
\cite{Kir1} and the Potts glass
\cite{Kir2} showed that this dynamical temperature is above that
predicted by the statics.  Recent studies of the off-equilibrium
dynamics of the $p$-spherical spin-glass model by Cugliandolo and
Kurchan have shown that the energy of the dynamics in the low
temperature phase, below the dynamical transition, is higher than that
predicted by the statics
\cite{CUKU}. As the dynamical transition temperature is approached the 
off-equilibrium dynamics slows down and aging effects start to appear.
Similar aging phenomena have been found in the context of random
manifolds \cite{Metw}.

In some cases the glass transition for the models without disorder can
be also predicted using the replica approach. A concrete test of all
these theoretical results for a deterministic model and its disordered
version (defined as the disordered model which has the same high-$T$
expansion as the deterministic model) has been performed very recently
for the sine model \cite{II} (hereafter referred as paper II). In this
case, the dynamical transition can be exactly computed and compared to
the numerical results. We stress on the fact that this dynamical
transition in the context of disordered systems corresponds to the glass
transition for the deterministic case.

The purpose of this work is to show how several numerical techniques in
spin glasses can be used for the determination of the glass transition
temperature for deterministic models. Because this glassy transition is,
as we have already indicated, purely dynamical this will be also the
main spirit of the techniques we will use. Now the reader will realize
that the main advantage of the study of deterministic models relies on
the fact that one does not need to average over different realizations
of disorder.  Because of the dynamical nature of the transition one
should average over different initial conditions. Anyway, comparing to
the spin-glass case we have eliminated one source of strong
fluctuations. In this work we will concentrate in the case of low
autocorrelation binary sequences. These models have received a lot of
attention very recently because they are the simplest prototype of
ordered systems with a very complex energy landscape. We adress the
reader to the differents works in this subject. Migliorini has performed
extensive numerical simulations using the tempering method \cite{Mi} and
Krauth and Mezard \cite{KrMe} and Krauth and Pluchery \cite{KrPl} have
applied a modified version of BKL algorithm which allows to investigate
dynamical properties for very large times.

The work is divides as follows. Section 2 introduces the low
autocorrelation models we will investigate and presents the main
theoretical results in this case. Section 3 is devoted to the study of
different thermodynamical quantities during an annealing cooling process
which clearly display the existence of the glassy transition. Section 4 is the
main nucleus of our work which is the study of off-equilibrium
properties of this models and particularly of aging.  The discontinuous
nature of the glass transition will enable us the use of this property
for an accurate prediction of the glass temperature.

\section{Low autocorrelation models}

By low autocorrelation models we denote a large class of deterministic
one dimensional models with long range interactions. In this work we
have focused our interest on the periodic and open models (so called
depending on the type of boundary conditions). These models have their
own interest as optimization problems in the field of communication
systems.  Let us suppose a one dimensional chain of Ising spins $\lbrace
\s_i;i=1,N \rbrace$ which can take the values $\pm 1$ and
the following Hamiltonian
\be
H=\frac{1}{N}\sum_{k=1}^N\,C_k^{2\nu} 
\label{eqH}
\ee
where the $C_k$ are correlation functions which connect spins at
distance $k$. The case $\nu=1$ is the problem on is generally interested
but nothing prevents of considering different models for a generic value of
$\nu$. For the periodic model we have,
\be
C_k=\sum_{i=1}^{N}\s_i\s_{i+k}
\label{periodic}
\ee
and in case of the open model,
\be
C_k=\sum_{i=1}^{N-k}\s_i\s_{i+k}
\label{open}
\ee
so in this case there is not translational symmetry in the model.

It was shown by Golay \cite{GOLAY} and after by Bernasconi \cite{BERNAS}
that one could approximate the thermodynamics of the open model by
supposing that the different correlation functions $C_k$ are
uncorrelated gaussian distributed random variables. This is the
Golay-Bernasconi (GB) approximation and predicts the existence of a
phase transition at a low temperature where the entropy vanishes. The
same conclusion is valid for the periodic model where one expects the
existence of a phase transition at low temperatures.

The interested reader can find most of the results of this section for
the periodic model in reference (I). Now we will recall some of the main
results obtained in that work. In case of the periodic model it can be
shown that for prime values of $N$ of the type ($4k+3$), $k$ being an
integer, there exists an explicit ground state of finite global energy
(and energy per spin zero in the thermodynamic limit because
eq.(\ref{eqH}) has to be normalized by $N$).  This construction does not
exist in the open case.  This ground state has a very low entropy up to
a finite temperature where the entropy experiences a sudden jump. This
finite temperature is the crystallization transition.  Starting at zero
temperature from the ground state configuration and slowly increasing
the temperature the entropy increases also very slowly (remaining always
very close to zero).  At the cristallization temperature the entropy
jumps to a finite value and the system enters into the high-temperature
regime. To account for this situation sometimes it is said that the
phase space has {\em 'golf course'} like properties. As regards to the
dynamical behaviour of these models, the existence of a ground state of
very low energy is of no relevance because we are interested
in the behaviour of large systems.  In fact, during a usual dynamical
relaxation process, the system is unable to find the ground state because
this state has very low entropy. In this situation, the particular
mathematical features of the selected number $N$ (prime of the type
$4k+3$ or not) are irrelevant. As has been shown in (I) the
high-temperature phase of the periodic model can be exactly solved.  Due
to the translational symmetry of the model one can write the Hamiltonian
in terms of the Fourier space components

\be 
B(p)=\frac{1}{\sqrt{N}}\sum_{j=1}^{N} \exp(\frac{2\pi i j
p}{N}) \s_j 
\label{Bp} 
\ee 
where $i$ stands for the imaginary unit. The Hamiltonian
eq.(\ref{periodic}) now reads

\be 
H=\frac{2}{N}\sum_{p=1}^{\frac{N}{2}} |B(p)|^4 \,\,.
\label{ham} 
\ee

Because the $\s_i$ are real functions (i.e.,
$B(p)=\overline{B(-p)}$) half of the Fourier components can be
neglected. 
Writing the Hamiltonian in the Fourier space one can show that only
certain kind of connected diagrams contribute to the free energy
allowing for a Hartree-Fock resummation of the full series. In another
way one can demonstrate, by introducing in eq.(\ref{Bp}) a generic
unitary matrix, that the replica approach can be used to find the free
energy of the model. In the replica symmetric approximation one recovers
the Hartree-Fock resummation. For our purposes it is important to note that the
free energy of the periodic model is given by
\be
f=\frac{1}{\beta}\,\log\int_0^{\infty}\,r\exp(-\beta
r^4-\mu r^2)dr~-\frac{1}{\beta}\log (2)-1
\label{free}
\ee
where the value of $\mu$ is determined by the equation
\be
\int_0^{\infty} r^3\,\exp(-\beta r^4-\mu r^2)\,dr=
\int_0^{\infty} r \,\exp(-\beta r^4-\mu r^2)\,dr
\label{c2}
\ee
The last condition corresponds to the closure condition
\be
\sum_{p=1}^{\frac{N}{2}}\,\langle |B_p|^2\rangle\,=1\,\,,
\label{BB}
\ee
the internal energy is given by

\be 
e=\frac{\partial\beta
f}{\partial\beta}=\sum_{p=1}^{\frac{N}{2}}\langle |B_p|^4\rangle -1
\label{UU} 
\ee

and the mean values $\langle ...\rangle$ are evaluated using the effective
Hamiltonian 
\be
{\cal H}(\lbrace B_p\rbrace)=-\beta \sum_p |B_p|^4-\mu \sum_p |B_p|^2\,\,.
\label{eff}
\ee
The integration variables $B_p$ are complex variables and the mean
values $\langle...\rangle$ are obtained integrating over the real and
imaginary part of the $B_p$.

These expressions are valid for the periodic model down to the
temperature at which the entropy vanishes, which is $T_{RSB}^C\sim 0.1$
(superindex $C$ means for periodic model and $O$ for the open case).
This result is surprisingly close to that given by the GB approximation.
We indicate that temperature with the subindex $RSB$ because at that
temperature replica symmetry is broken.  Below $T_{RSB}^C$ the entropy
is nearly zero and the energy is constant, a situation indeed very
similar to that of spin-glasses with one step of replica symmetry
breaking \cite{DERRIDA,GROMEZ}.  Obviously the previous expression
eq.(\ref{free}) is not valid for the open model for which a
high-temperature resummation is still lacking. Anyway, using the Golay
approximation we can estimate this transition to be close to
$T_{RSB}^O\sim 0.047$.

As we will see in the following sections the transition $T_{RSB}$ for
the periodic and for the open model are not the true ${\it glass~
transition}$. As discussed in the introduction, the true glass
transition corresponds to the transition where dynamical stability is
lost (ie. the temperature given by the marginality condition) and it can
be several times larger than the corresponding transition given by the
statics. This result was already known in spin-glasses but it is new to
know that this result also applies in case of deterministic models. The
following sections are devoted to the numerical determination of this
transition in the open and periodic case. We will see also how
off-equilibrium phenomena and particularly the property of aging yield a
very nice preciser to determine this glass transition temperature.
Regarding the dynamical behaviour of both models we can advance that the
main conclusions will be the same for the open and the periodic case.
Because the open model has historically received more attention than the
periodic version we will present more dynamical results in the former
case.

\section{A first determination of the glass 
transition}

The main property of the glass transition in low autocorrelation models
regards the first order nature of this dynamical transition. From the
thermodynamic point of view this transition is second order. So, for
instance, the energy and the entropy are continuous while the specific
heat experiences a jump. Because the transition is purely dynamical, this
implies a transition for the correlation and response functions.  In
this section we will explore the behaviour of thermodynamic quantities
leaving the discontinuous feature of the order parameter to the next
section.  For the periodic and open models we have done the same kind of
studies. In fact, we have discovered that they are strongly similar
except for the fact that the periodic model is analytically solved in
the high $T$ phase and displays an explicit ground state for chain
lengths $N$ such that $N$ is prime and of the type $4k+3$, $k$ being an
integer.  

Starting from a random initial configuration in the high-temperature
region we have decreased progressively the temperature in a Monte Carlo
annealing. We have simulated several sizes up to $N=1000$ (because it is
a long range problem the number of bit operations in a Monte Carlo
updating procedure grows very fast with the size of the system). We have
also tested that finite-size corrections are negligible and different
initial conditions give the same result. As said in the introduction, we
have now only one realization of disorder on which we have to do
simulations.  We have computed the main thermodynamical observables like
the energy, magnetization and their associated dissipative quantities
like specific heat and magnetic susceptibility.  The behaviour of the
energy are shown in figures 1 and 2 respectively for the periodic and
open model. The energy decreases down to a certain temperature where it
remains constant. This is much similar to what happens in the REM
\cite{DERRIDA}. The dashed line in figures 1 and 2 corresponds to the GB
approximation and the continuous line (only for the periodic model)
corresponds to the correct high-temperature prediction eq.(\ref{free})
which is in agreement with the data. As has been already commented, the
glass transition is higher than the static transition (close to $0.1$ in
the periodic model). Figure 1 shows where the entropy of the
high-temperature expression of eq.(\ref{free}) vanishes. Curiously it
does at the same temperature as that given by the GB approximation. We
have no explanation for this result.  If this were true also in the
opened case one would be tempted to state that the GB approximation is
enough to predict the static transition.  Figures 3 and 4 show the
behaviour of the specific heat for the periodic and open model
respectively. Also in this cases we plot the results for the GB
approximation and, in case of the periodic model, we plot also the
high-temperature prediction eq.(\ref{free}). In both cases we observe a
discontinuous jump of the specific heat. It appears at a temperature
$T_G^C\sim 0.45$ for the periodic model and $T_G^O\sim 0.2$ for the open
case.  We have to note that this energy and specific heat in the low $T$
phase are purely dynamical. Anyway they satisfy fluctuation dissipation
theorems like the relation $C=\frac{\partial e}{\partial T}$ where $C$
is the specific heat and $e$ is the internal energy.

We have also measured the magnetization and its fluctuations. The global
magnetization is zero above $T_G$ and below that temperature remains
stacked to a certain small non zero value (of the order of the standard
mean deviation $1/\sqrt{N}$). Valuable information can be obtained from
its fluctuations like the linear susceptibility and the Binder
parameter. If $P(M)$ is the probability distribution of the
magnetization, we expect it will be a Gaussian at very large
temperatures and become more and more flattened as the glass transition
is approached. We are going to show that this is really the case and that
fluctuations are very large even if we stay at high temperatures. In plain
words, the linear susceptibility and the Binder parameter are the
variance and the curtosis of the probability distribution $P(M)$. The
linear susceptibility is given by

\be 
\chi=\beta (\langle M^2\rangle-\langle
M\rangle^2) 
\label{eqchi} 
\ee 

where $M$ is the global magnetization and we recall the fact that one
factor $N$ has been absorbed in the temperature in order to have an
appropriate thermodynamic limit.  The Binder parameter is given by

\be 
g=\frac{1}{2} (3-\frac{\langle M^4\rangle}{\langle
M^2\rangle^2}) 
\ee

Now we would like to compute approximately this quantities in the high
$T$ phase above the glass transition. From eq.(\ref{ham}) we observe
that the Hamiltonian is the sum of $\frac{N}{2}$ Fourier components
$B(p)$.  We can suppose that these Fourier components are independent at
least in the high $T$ phase (in some sense this is the original idea of
Golay in order to resumme the high $T$ series). One can soon realize
that this approximation has to fail because the total number of Fourier
components is too large (it diverges with $N$). But this is the easiest
approximation one can do. In order to reach the correct expression it
should be necessary to solve the low autocorrelation models in a
magnetic field. Within this approximation and using the Hamiltonian
eq.(\ref{ham}) we observe that the zero momentum term $|B(0)|^4$
corresponds to the fourth power of the magnetization. The only
difference between the magnetization and the $B_p$ is that these last
Fourier components are complex while the magnetization is real.
According to eq.(\ref{eff}) the effective probability distribution of
the magnetization is given by

\be
P(M)\sim \exp (-\beta M^4- \mu M^2)
\label{eqPM}
\ee

We immediately observe that only at infinite temperature the probability
distribution will be a Gaussian and at finite $\beta$ non Gaussian
corrections can be very strong (the same discussion is valid for any
Fourier component $B_p$). This result was numerically observed by
Migliorini studying the local field distribution \cite{Mi} in the open
model. Using this approximation and equation (\ref{BB}) for the periodic
model one gets
\be
\chi=\beta
\label{chi}
\ee

for the linear susceptibility of the periodic model. Figures 5 and 6 show
the linear susceptibility obtained during an annealing process. Figure 5
also shows the prediction eq.(\ref{chi}) for the periodic model.  The
values obtained for the glass transition from the discontinuity of the
linear susceptibility agree with those obtained measuring the specific
heat (figures 3 and 4).

In case of the Binder parameter we use eq.(\ref{UU}) and we can obtain
it in terms of the internal energy (now one has to be a little bit
careful and realize that the integral of the fourth power of the
magnetization, which is a real variable, over the probability
distribution eq.(\ref{eqPM}) is $\frac{3}{2}$ times the integral of the
fourth power of any complex Fourier component $B_p$ over the effective
Hamiltonian eq.(\ref{eff})). One gets the result,

\be
g=\frac{3}{4}(1-e)
\label{g}
\ee

We show in figure 7 the behaviour of the Binder parameter associated to
the magnetization for the periodic model (Monte Carlo results are also
shown for the open case). It is shown up to $T=2$ (five times the
predicted glass temperature of the periodic model). For very large
temperatures the Binder parameter should vanish because the
magnetization distribution becomes a Gaussian. In our case it decays
very slowly to zero which indicates that well above the glass
temperature fluctuations in the magnetization are large. Also from
figure 7 we can observe a a jump for the Binder parameter at the glass
transition to a value close to $1$.  One comment about the high value of
$g$ above $T_G$ is now appropriate.  This large value of the curtosis
parameter means that the probability distribution of the magnetization
is far from being a Gaussian. It is a symmetric distribution very flat
close to $M=0$ and possibly with two peaks symmetrically distributed. As
we will see in the next section, this result has strong implications for
the dynamics. We expect that well above $T_G$ the spin-spin correlation
function $\langle
\s(t_0)\s(t)\rangle$ decays to zero very fast but 
the system can preserve a certain memory of the configuration at time
$t_0$. In fact, if the $P(M)$ is so much flattened around $M=0$, the system
can need a very large time to reach configurations completely uncorrelated
from the memorized configuration at $t_0$.

Let us summarize the results of this section. Doing annealings, starting
from large temperatures down to the low $T$ region, we observe a glass
transition where the energy freezes and fluctuations vanish. This
temperature is several times larger than that predicted by the statics
and this is related to the peculiar structure of the high energy
metastable states which the systems explores during the relaxation.
More concretely, we have learned that the glassy temperature occurs at
$T_G^C\simeq 0.45$ for the periodic model and $T_G^O\simeq 0.2$ for the
open case. In the next section we will confirm this results by studying
the off-equilibrium dynamics of these models. In particular, aging
phenomena will appear as a nice method to determine the glass
transition.

\section{Aging and the first order nature of the dynamical transition}

As we said in the last section this transition is of first order nature
in the dynamical order parameter. In principle the dynamics is described
by the two-time correlation functions $C(t_1,t_2)$ and the response 
functions $G(t_1,t_2)$. They
are defined as usual by:
\bea
C(t_1,t_2)&=&\langle \s_i(t_1)\s_i(t_2) \rangle\\
G(t_1,t_2)&=&\frac{\delta \langle \s_i(t_1)\rangle}{\delta
h(t_2)}~~~~t_2 < t_1\\
\eea

where $\langle...\rangle$ is the usual time average over different
noise realizations in the dynamics and $h(t_2)$ is the magnetic field
applied to the system at the time $t_2$. We have
performed discrete Monte Carlo dynamics which we expect to give similar
results as well as a usual Langevin dynamical process.

In the high-temperature regime, above the glass transition, we expect
that the correlation and the response functions are related one to the
other by the fluctuation-dissipation theorem. Also in this high $T$
region the correlation and the response functions satisfy the
time-homogeneity hypothesis, i.e. the functions $C(t_1,t_2)$ and
$G(t_1,t_2)$ depend only on the time difference $t_1-t_2$. Both
functions decay very fast in time.

Below the glass transition the time behaviour of the correlation and
response function drastically change and, for instance, time
correlations decay very slowly in time. In this low $T$ regime the time
homogeneity hypothesis is lost and strong aging effects start to appear.
Then the decay of the correlation functions depends on the previous
history of the system. More concretely, it depends on the time $t_1$ at
which the spin configuration is memorized (in case of the
correlation functions) or on the time $t_2$ at which the magnetic field
is switched off (in case of the response function).

For simplicity reasons we have focused our
research on the two-time correlation function (one could also perform
aging experiments measuring the remanent magnetization). In this case we have
measured the time-time correlation function between the spins
configuration at time $t_w$ and the configuration at time $t_w+t$,

\be
C(t_w,t_w+t)=\frac{1}{N}\,\sum_{i=1}^{N}\,\s_i(t_w)\,\s_i(t_w+t)
\label{ctt}
\ee

Above the glass transition temperature we expect time homogeneity
applies (this means that $C(t_w,t_w+t)$ only depends on $t$) and time
correlation functions should decay very fast to zero. The following
condition holds
\be
\lim_{t_w\to\infty}\,C(t_w,2t_w)=0 \,\,.
\label{eq1}
\ee

Just below $T_G$ the correlation function decays very slowly in time to
a finite value $q_1$. This finite value $q_1$ is positive and smaller
than the static Edwards-Anderson order parameter at the static
transition point. This value $q_1$ is zero above $T_G$ and is very close
to 1 just below $T_G$ and increases as the temperature decreases
(linearly with $T$ at low temperatures). We have to call the attention
of the reader to the fact that this value is physically related to the
local order parameter associated to the metastable states and this is
smaller than the local overlap associated to the true equilibrium
configurations (the static Edwards-Anderson order parameter). The
procedure in order to determine the value of $q_1$ has been applied
recently to a particular deterministic model (see (II)) and corresponds
to the replica order parameter within the same block at one step replica
of replica symmetry breaking. This is evaluated at the dynamical
transition point where the free energy is maximized according to the
marginality condition. More precisely, we can write (for an infinite
system)

\be
\lim_{t_w\to\infty}\,C(t_w,2t_w)=q_1
\label{eq2}
\ee
where $q_1$ depends on the temperature. For low autocorrelation models
we know the value of $q_1$ is very close to 1 (for instance, this is the
greatest difference between $p$-spin glasses \cite{GARDNER} or Potts
glasses \cite{EMILIO} and low autocorrelation models, the last ones show
a very large discontinuity in the value of $q_1$). Because the value of
$q_1$ jumps from zero above $T_G$ to a finite value below $T_G$ the
transition is of a discontinuous type.  Before showing our dynamical
results in case of low autocorrelation binary sequences we would like to
note that, as regards to the dynamical experiments, deterministic models
are much more suited than disordered models. Because our model is
ordered, we do not need to save the realization of the random couplings.
The number of random couplings, in case of a long range model, can be
very large and this sets a limit on the maximal size one is able to
memoryze in the computer. The major part of the numerical results we
will show correspond to $N=5000$ in both models (open as well as in the
periodic case).

The existence of aging is one of the most outstanding features of spin
glasses \cite{AGING}. Now we are going to show that also low
autocorrelation models also exhibit these phenomena just below the glass
transition. Because the results we have obtained for the periodic and
the open case are very similar, we will present in some cases the
results only for the open case. Figure 8 shows the correlation function
eq.(\ref{ctt}) for the open model, for different values of the waiting
time above the glass transition $T_G$ (as estimated in the previous
section).  The data in this case corresponds to a temperature $T=0.45$.
This figure shows that above the glass transition the aging effects are
absent (i.e., the correlation functions do not depend on the value of
$t_w$).  Also, all correlation functions decay very fast with the time.
Surprisingly (as shown in the figure 8) they do not decay always to
zero. In some cases, they decay to a small finite value (for the
suspicious reader we will note that this value is fairly large than the
standard deviation $\frac{1}{\sqrt{N}})$. This means that, well above
the glass transition, the system preserves a small temporal correlation
with previous configurations. As discussed in the previous section,
this is strongly related to the non gaussianity of the fluctuations (for
instance, this was shown in the case of the magnetization). This
behaviour is far from being paramagnetic. It is not clear to us what is
the real dynamical nature of this high-$T$ region.

As soon as we go below $T_G$ the dynamics slows down dramatically. The
system remains trapped in metastable states and it takes a very long
time for the system to overcome the barriers and explore new
configurations. This is clearly seen in the results of figure 9 where we
show the correlation function below the glass transition at $T=0.1$ for
one realization of the noise for the open model. Aging effects are
present and we expect correlation functions to depend mainly on the
ratio $t/t_w$, for large enough values of $t_w$. Some comments are now in
order. As shown in figure 9 the correlation function stays very close to
1 during a time of order of $10^4$ Monte Carlo steps for all different
waiting times.  This is because for enough low temperatures the system
is able to surmount only a few number of states and the shape of the
correlation function is strongly dependent on the noise realization. To
get smooth correlation functions one should average over a very large
number of trajectories and this demands for a lot of computer time. From
this considerations it emerges that a scaling law of the type

\be 
C(t_w,t+t_w)\sim f(t/t_w) 
\label{scale} 
\ee 

is very difficult to observe in a small number of decades of time.  This
scaling law has been obtained by Bouchaud in his phenomenological
approach to the off-equilibrium dynamics \cite{BOU1}. Cugliandolo and
Kurchan \cite{CUKU} have explicitely shown that this is indeed a
solution of the off-equilibrium equations in case of the $p$-spin
spherical spin-glass model and the Potts model \cite{POCU}.  These
models do have a spin-glass phase with one step of replica symmetry
breaking. It is reasonable to suppose that the scaling law
eq.(\ref{scale}) also applies in case of low autocorrelation models for
which a REM-like transition describes well the low $T$ behaviour.  We
should also note that the dynamical behaviour we are observing in these
models is strongly different from the dynamical relaxation of the SK model
\cite{CUKURI} or short-ranged Ising spin glasses \cite{VARI}. In that
case, one does not has a first order dynamical transition and the free
energy landscape is not so rough. The system is not trapped in the
metastable states and correlation functions decay to zero smoothly
without apparent jumps \cite{CUKU2}. When a strong metastability is
present (like in low autocorrelation models) new numerical techniques
like those recently developed by Krauth and Pluchery
\cite{KrPl} and Krauth and Mezard \cite{KrMe} are very useful.  If one
wants to observe smooth aging in a reasonable scale of time, it is
mandatory to go to higher temperatures.  Precisely, at the glass
temperature, we expect that the system will display nice aging and the
scaling law eq.(\ref{scale}) will be satisfied for enough large sizes.
This is shown in figure 10 where we have measured the aging at a
temperature $T_G\sim 0.19$ for the open model and a very large size
$N=10000$. The inset of figure 10 shows the scaling law
eq.(\ref{scale}).

Now we want to show how aging allows for a nice confirmation of the
first order nature of the glass transition. This is one of the main
results of this work. Because the nature of this glass transition is
purely dynamical we can use the relations eq.(\ref{eq1}) and
eq.(\ref{eq2}) in order to find the temperature at which the
discontinuity of the order parameter appears. A similar technique could
be used by coupling two replicas like has been done in case of the
$p$-spherical spin glass model \cite{KUPAVI}.  Nevertheless, we think
that our dynamical technique is more direct because we do not need the
introduction of an extra coupling parameter in the model.

We have computed the correlation function for different waiting times
$t_w$ and also different temperatures. Then, for each temperature, we
computed $C(t_w,2t_w)$ averaging the correlation function in a
logarithmic scale. We proceeded in this way in order to get smooth values of
the correlation $C(t_w,2t_w)$ as a function of the temperature and the
waiting time.  We have done this numerical analysis for different values
of $t_w=100,300,1000$ in the periodic model and $t_w=100,300,1000,3000$
for the open model.  Figures 11 and 12 show the results for the periodic
and open model respectively. From this data we can see clearly the
discontinuity because the predicted value of $q_1$ is very close to 1 just
below $T_G$.  

In order to obtain $T_G$ we have performed a usual finite-time scaling
analysis. To this end we have measured the relaxation curves above the
glass transition and also above the temperature at which finite-size
effects are negligible (approximately $T=0.25$ for the open model and
$T=0.55$ in the periodic case). Correlation functions decay
exponentially and one can estimate the relaxation time $\tau$ as a
function of $T$. In this range of temperatures we expect the correlation
time will diverge as a power law singularity of the type

\be
\tau\sim (T-T_G)^{-\gamma}
\label{tau}
\ee

where $\gamma$ is a dynamical exponent. We note that this kind of
divergence is typical also of disordered systems with long or
short-range interactions. In case of frustrated models without disorder
the situation can be different depending on the range of the
interaction. Low autocorrelation models are of the long-ranged type. It
is possible that for more realistic models of glasses the dynamics will
be much more complex and strongly different relaxation behaviours, like
the Arrhenius or the Vogel-Fulcher law, could take place. Now we want to
observe that usual critical dynamics works well in the case of low
autocorrelation models. This is not surprising if (as we have seen in
this work) glasses and spin glasses do have so much in common
\cite{SOULETIE}. We have fitted the correlation functions in the high
$T$ regime with a scaling law of the type

\be
c(t)\sim f(t/\tau)
\label{scaling}
\ee

where $\tau$ is given in eq.(\ref{tau}). The scaling behaviour is shown
in figures 13 and 14 for the open model and the periodic model
respectively. Good fits are obtained with $T_G\sim 0.21\pm0.02 $ and
$\gamma\sim 2\pm 0.5$ for the open model and $T_G\sim 0.43\pm 0.2$,
$\gamma\sim 2\pm 0.5$ for the periodic model. The scaling function
$f(t/\tau)$ is nearly an exponential in both cases. The exponent
$\gamma$ is the equivalent of the product of exponents $z\nu$ for usual
critical dynamics and it is certainly much lower than known values in
realistic glasses (tipically these are of order 10, see
\cite{SOULETIE}). As we have already indicated, low autocorrelation models
are long-ranged models. Realistic glasses are not of the long range type
and it could well be that the exponent $\gamma$ increases as the
dimensionality decreases. This happens in case of Ising spin glasses
where the product $z\nu$ ranges from $2$ in mean-field theory to $6$ in
three dimensions \cite{OGIELSKY} (in case there is a true phase
transition \cite{MAPARI})

\section{Conclusions}

Low autocorrelation models display a dynamical
behaviour very similar to disordered spin glasses. The reason for this
similarity is that these models (and more generally,
glasses) do have a broad distribution of higher free energy metastable
states like happens in case of spin glasses \cite{BrMo}. 

The feeling which emerges from recent studies by
several groups is that deterministic models display a glassy behaviour
of a purely dynamical nature. This glassy behaviour seems to be
associated to spin glass models with one step of replica symmetry
breaking \cite{GROMEZ}. In the case of models with an infinite number of
breakings like the SK Model \cite{SK} the situation is different
\cite{CUKU2}.

We have also seen that the open case and the
periodic case behave very similarly. We have studied the relaxation
of magnitudes like the internal energy, specific heat and magnetic
susceptibility. More interestingly, the Binder parameter associated to
the magnetization has a non Gaussian shape even for very large
temperatures above the glass transition. This result should apply very
probably also for any other Fourier component $B_p$ of the configuration
of the spins. 

According to this result we have seen that well above the glass
transition the dynamical correlation functions decay exponentially fast
to a small non zero value. The system is not fully ergodic because it
has some memory about the previous configurations it has visited. We
have given an explanation to this fact but it remains unsolved what is
the real nature of this high-$T$ phase.  Above the glass transition
temperature finite-time scaling analysis has revealed a good technique
in order to locate the transition and the dynamical exponents. For the
open and periodic models we obtain the equivalent of the product
exponents $z\nu$ of the critical dynamical theory. Values close to 2 are
obtained. Compared to experimental values obtained in case of real
glasses these are small. But this could be an artifact of the long-ranged
interactions of the low autocorrelation models.

We have also investigated the dynamics below the glass transition where
aging phenomena is present. This is one of the main features in spin
glasses. At the glass transition, where the effect of the
traps is not very strong, we have found that the scaling law
eq.(\ref{scale}) is well reproduced.

We conclude by saying that the techniques developed in this work are
very general and should be applicable to a large variety of systems
where disorder is not present. In particular we think that the behaviour
of $c(t_w,2t_w)$ as a function of the temperature is well suited in
order to characterize the transition. It remains to be understood to
what extent the results obtained in this work are generic for real
glasses and to what extent short-range interactions can modify the main
conclusions of this work. 

\section{Acknowledgements}

We are grateful to Leticia Cugliandolo, Jorge Kurchan, Enzo Marinari,
Marc Mezard and Giorgio Parisi for stimulating and continuous discussion
on these subjects.  One of us (F.R.) acknowledges the INFN for financial
support.

\vfill
\newpage
{\bf Figure caption}
\begin{itemize}

\item[Fig.~1] Energy of the periodic model versus temperature.
The continuous line is the high-temperature result eq.(\ref{free}). The
dashed line is the GB approximation. Simulation results are for $N=100$

\item[Fig.~2] Energy of the open model versus temperature. 
The dashed line is the GB approximation. Simulation results are for
$N=500$ (squares) and $N=1000$ (crosses)

\item[Fig.~3] Specific heat of the periodic model versus temperature.
The continuous line is the high-temperature result eq.(\ref{free}). The dashed
line is the GB approximation. Simulation results are for $N=100$
(squares) and $N=500$ (crosses).

\item[Fig.~4] Specific heat of the open model versus temperature. The dashed
line is the GB approximation. Simulation results are for 
$N=100$ (squares) and $N=500$ (crosses).

\item[Fig.~5] Magnetic susceptibility of the periodic model versus
temperature. The continuous line is the approximation eq.(\ref{chi}). 
Simulation results are for $N=100$ (squares) and $N=500$ (crosses).

\item[Fig.~6] Magnetic susceptibility of the open model versus
temperature. Simulation results are for $N=500$ (squares) and $N=1000$
(crosses).

\item[Fig.~7] Binder parameter of the periodic and open models 
versus temperature. The continuous line is the approximate
high-temperature result to the periodic case eq.(\ref{g}). Data is shown
for $N=100$ in the periodic model and $N=500$ in the open case. 

\item[Fig.~8] $C(t_w,t+t_w)$ for the open model for different values of
$t_w$ above the glass transition at $T=0.45$. The size is $N=5000$.

\item[Fig.~9] $C(t_w,t+t_w)$ for the open model for different values of
$t_w$ below the glass transition at $T=0.1$. The size is $N=5000$.

\item[Fig.~10] $C(t_w,t+t_w)$ for the open model for different values of
$t_w$ at the glass transition ($T_G\sim 0.19)$. The inset shows
the scaling law eq.(\ref{scale}). The size is $N=10000$.

\item[Fig.~11] $C(t_w,2t_w)$ for the periodic model for different values of
$t_w=30,100,300,1000$ as a function of the temperature. 

\item[Fig.~12] $C(t_w,2t_w)$ for the open model for different values of
$t_w=30,100,300,1000,3000$ as a function of the temperature. 

\item[Fig.~13] Finite-time scaling eq.(\ref{scaling}) for the periodic case.
Good scaling is obtained with $T_G\sim 0.43\pm $ and $\gamma\sim 2$.

\item[Fig.~14] Finite-time scaling eq.(\ref{scaling}) for the open case.
Good scaling is obtained with $T_G\sim 0.21\pm $ and $\gamma\sim 2$.

\end{itemize}
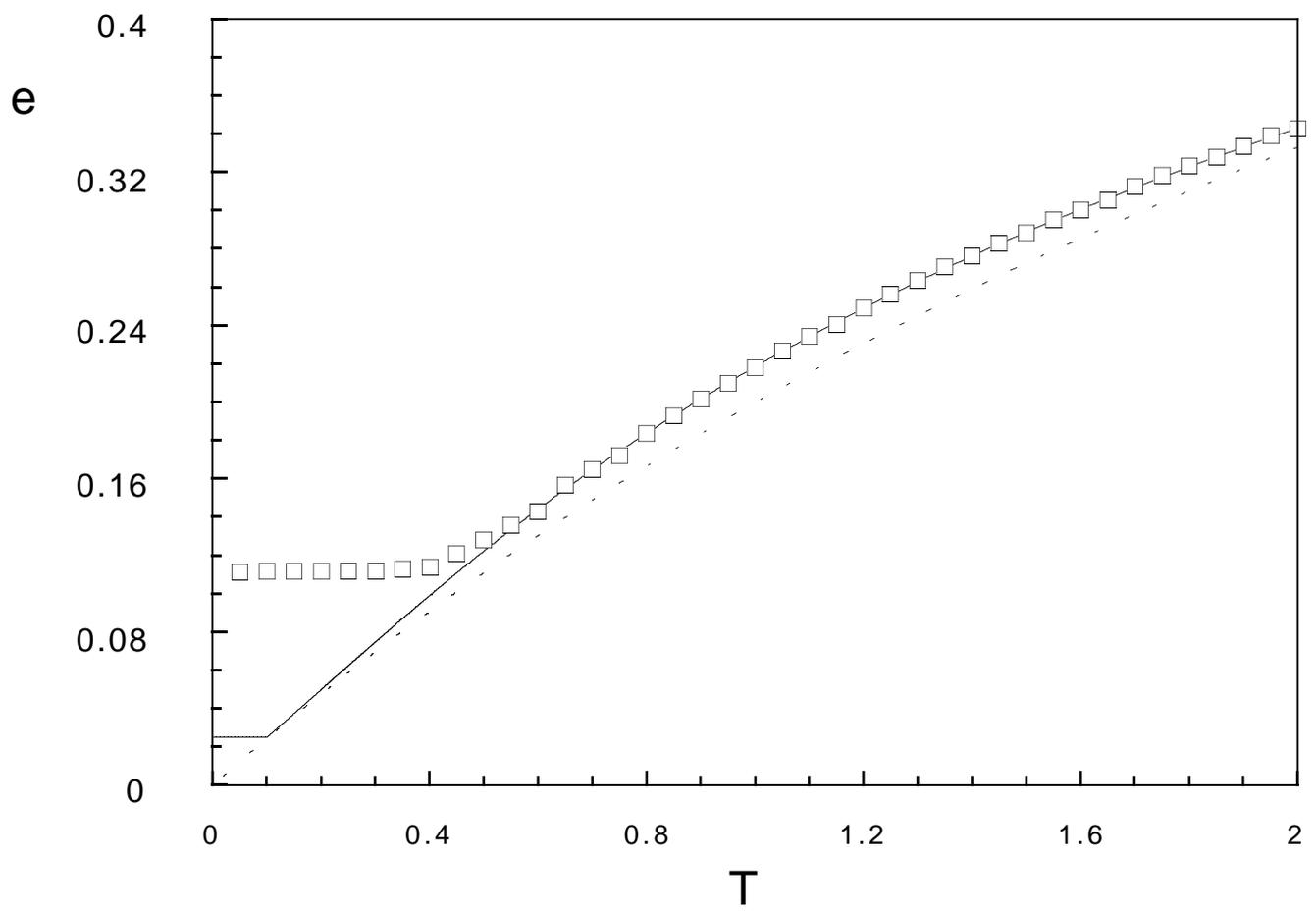
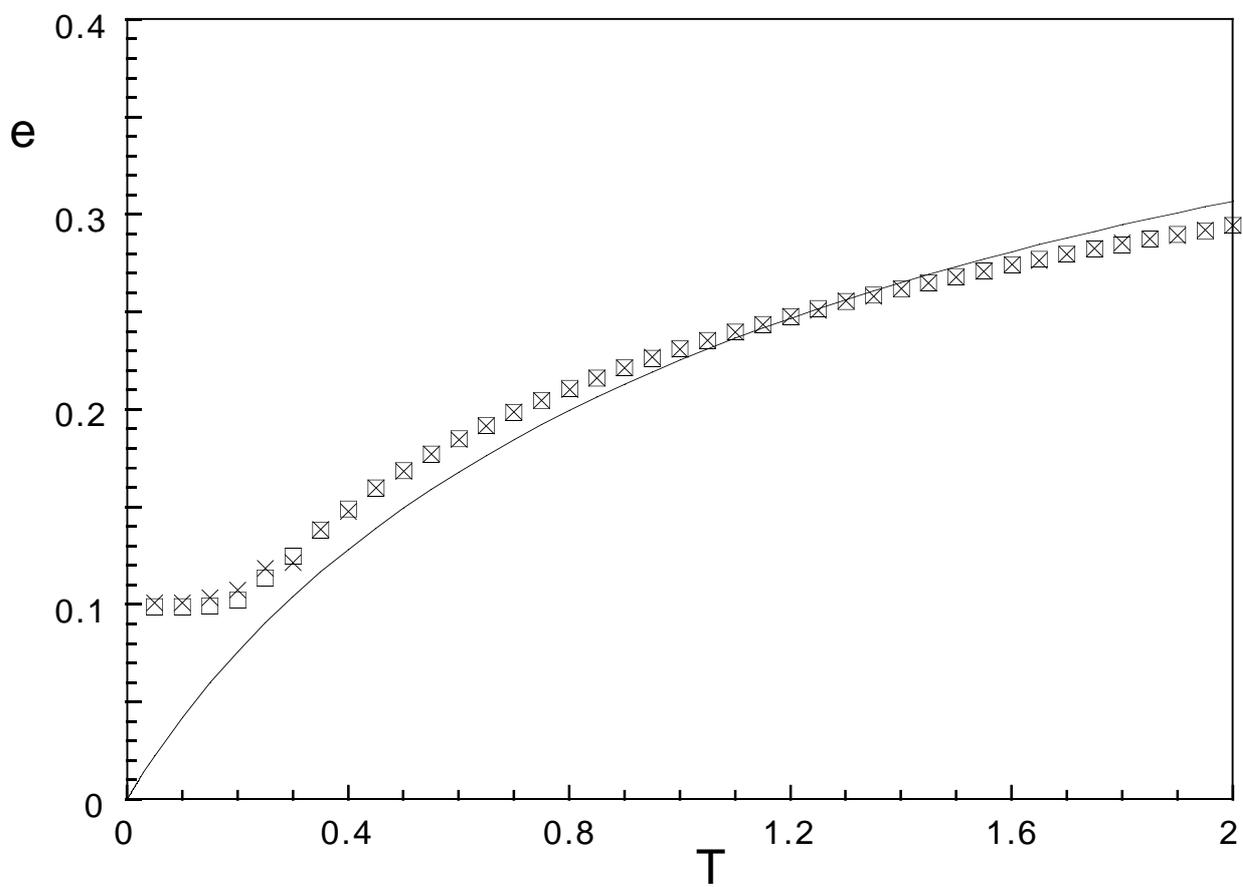
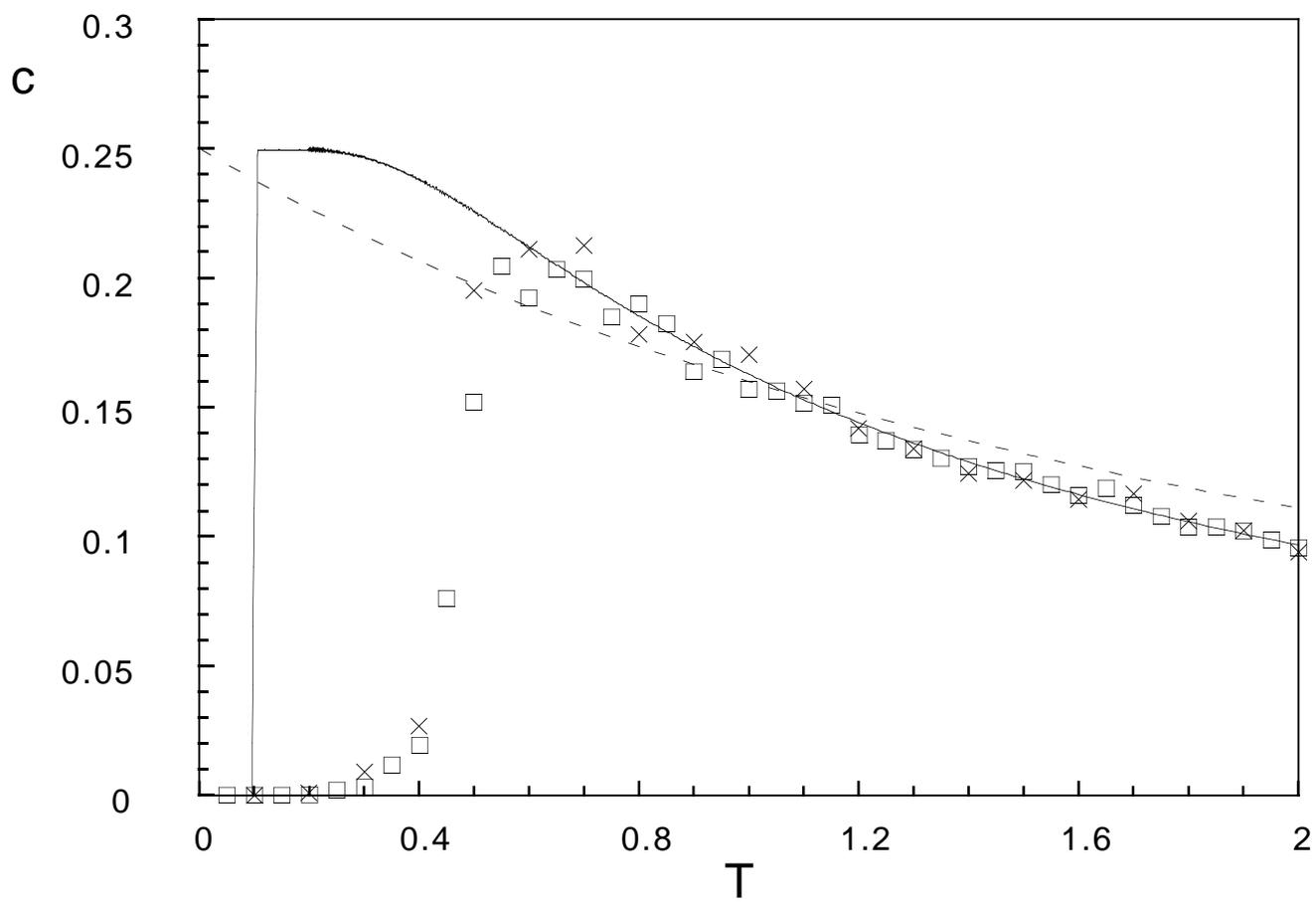
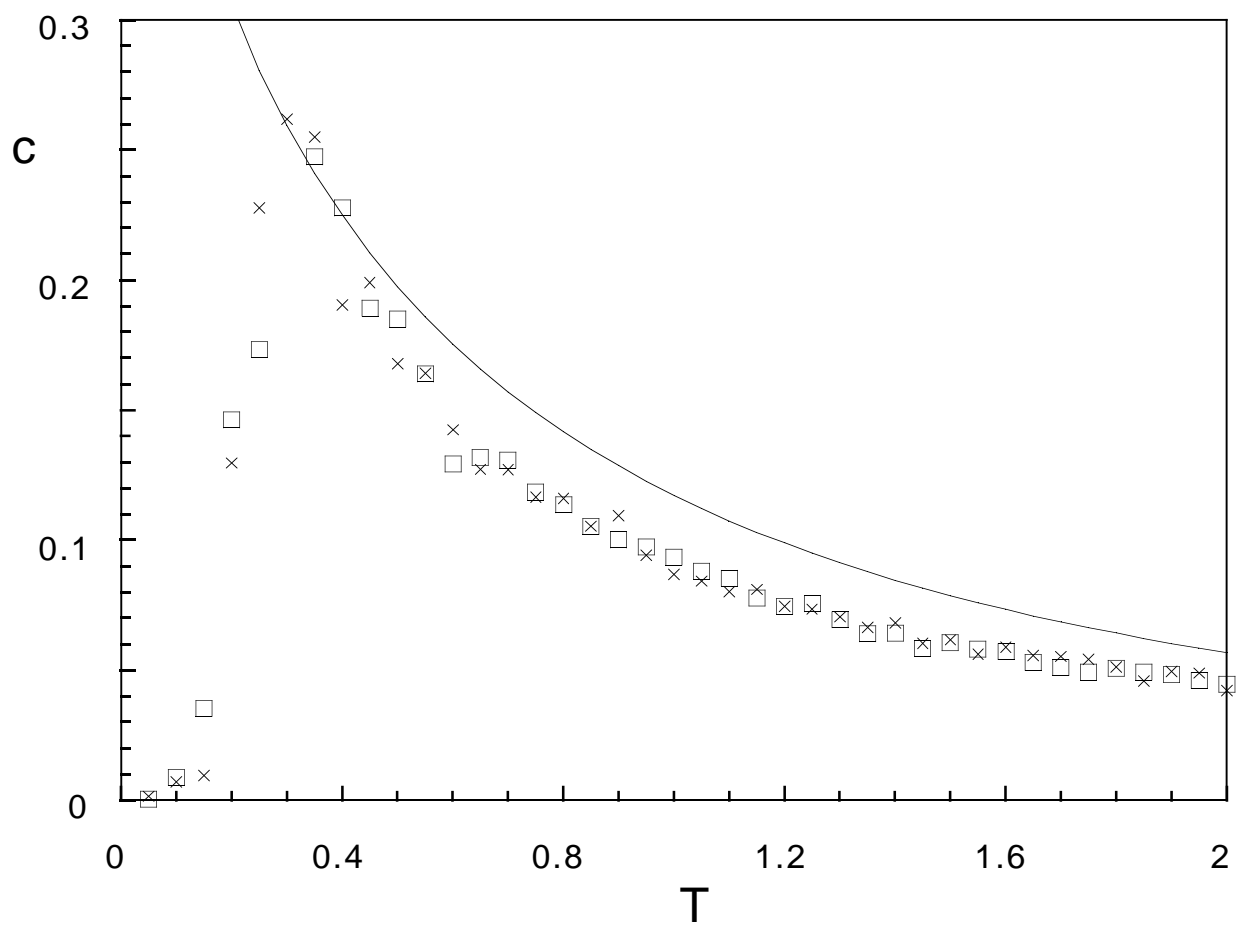
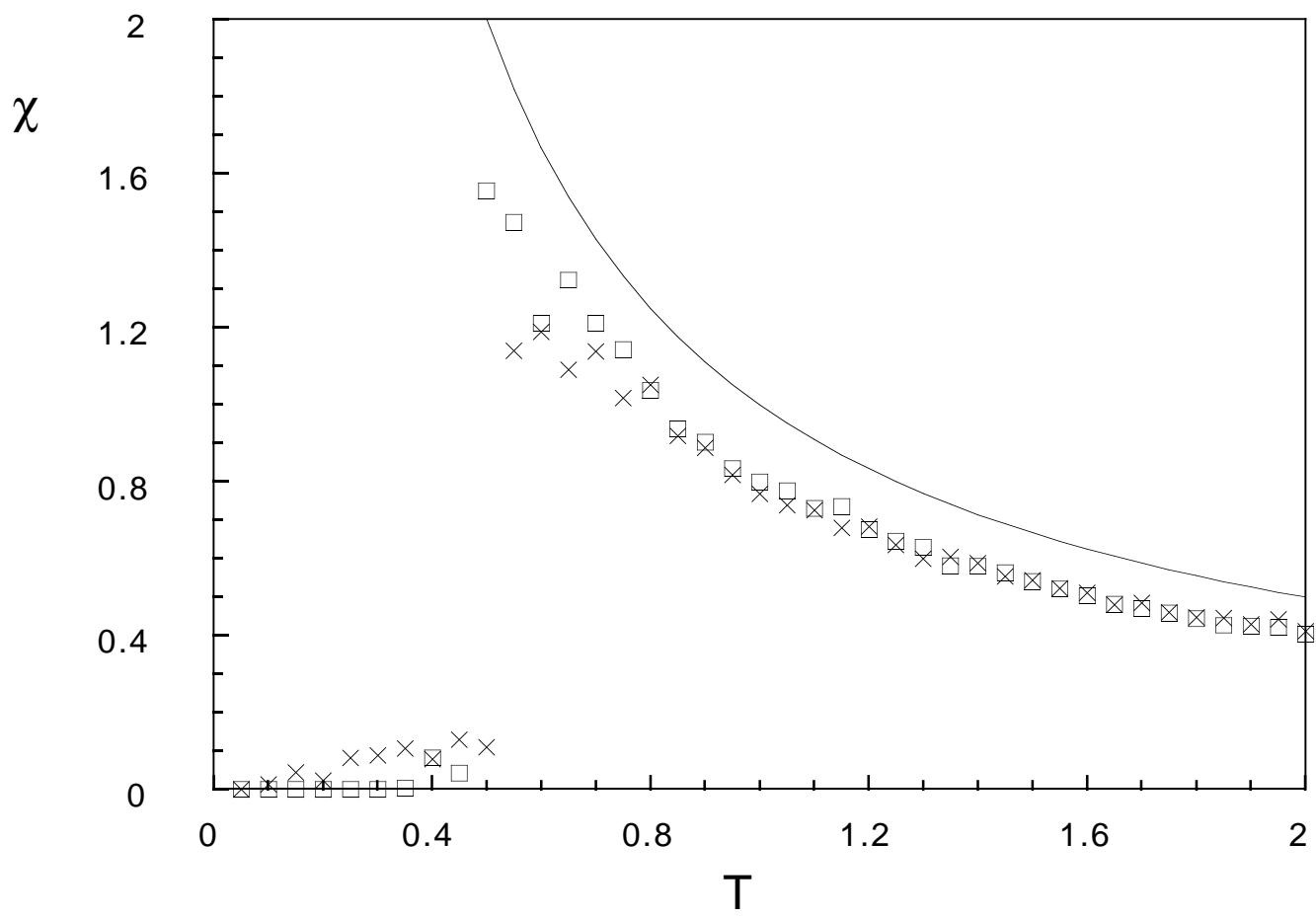
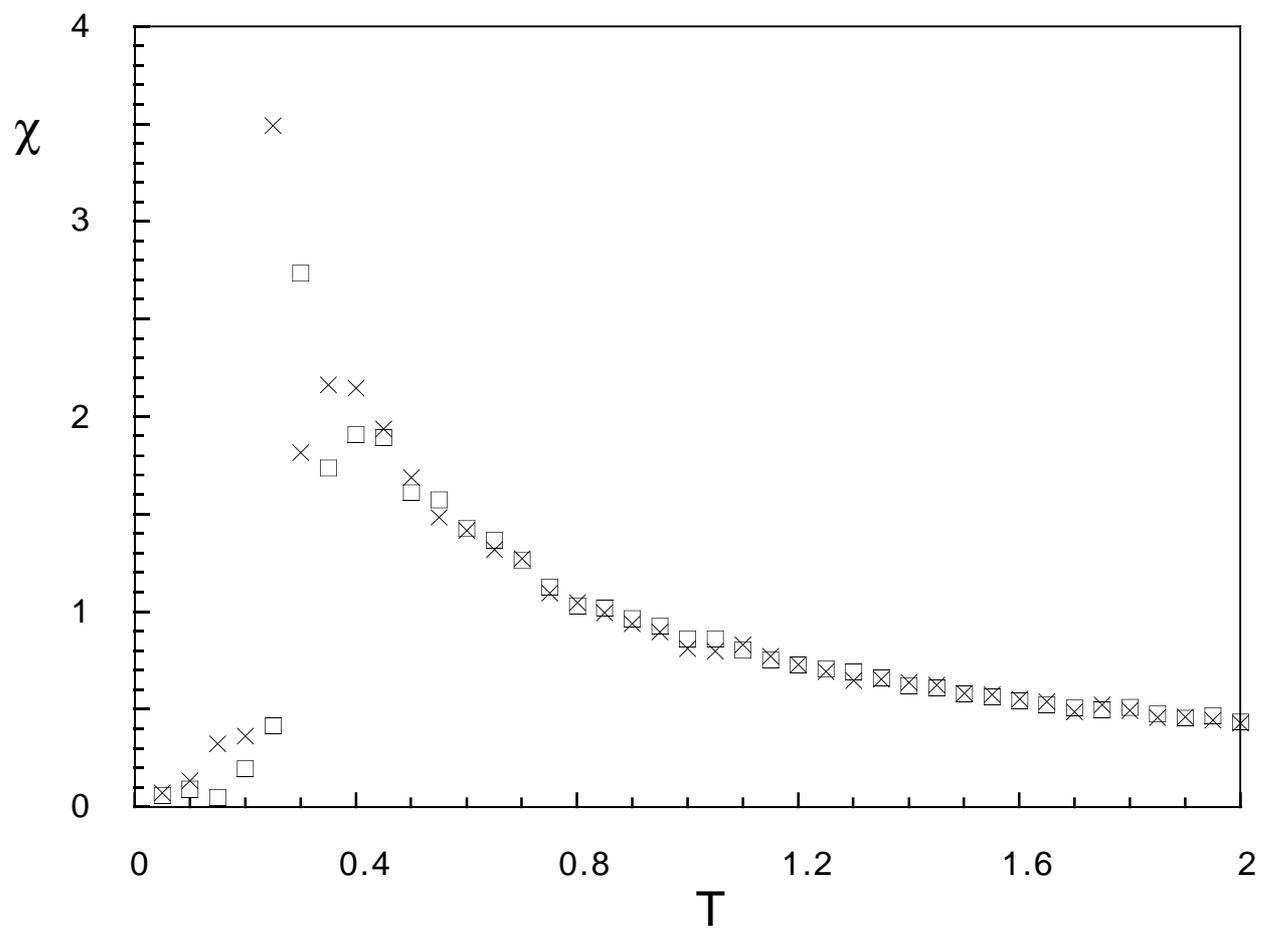
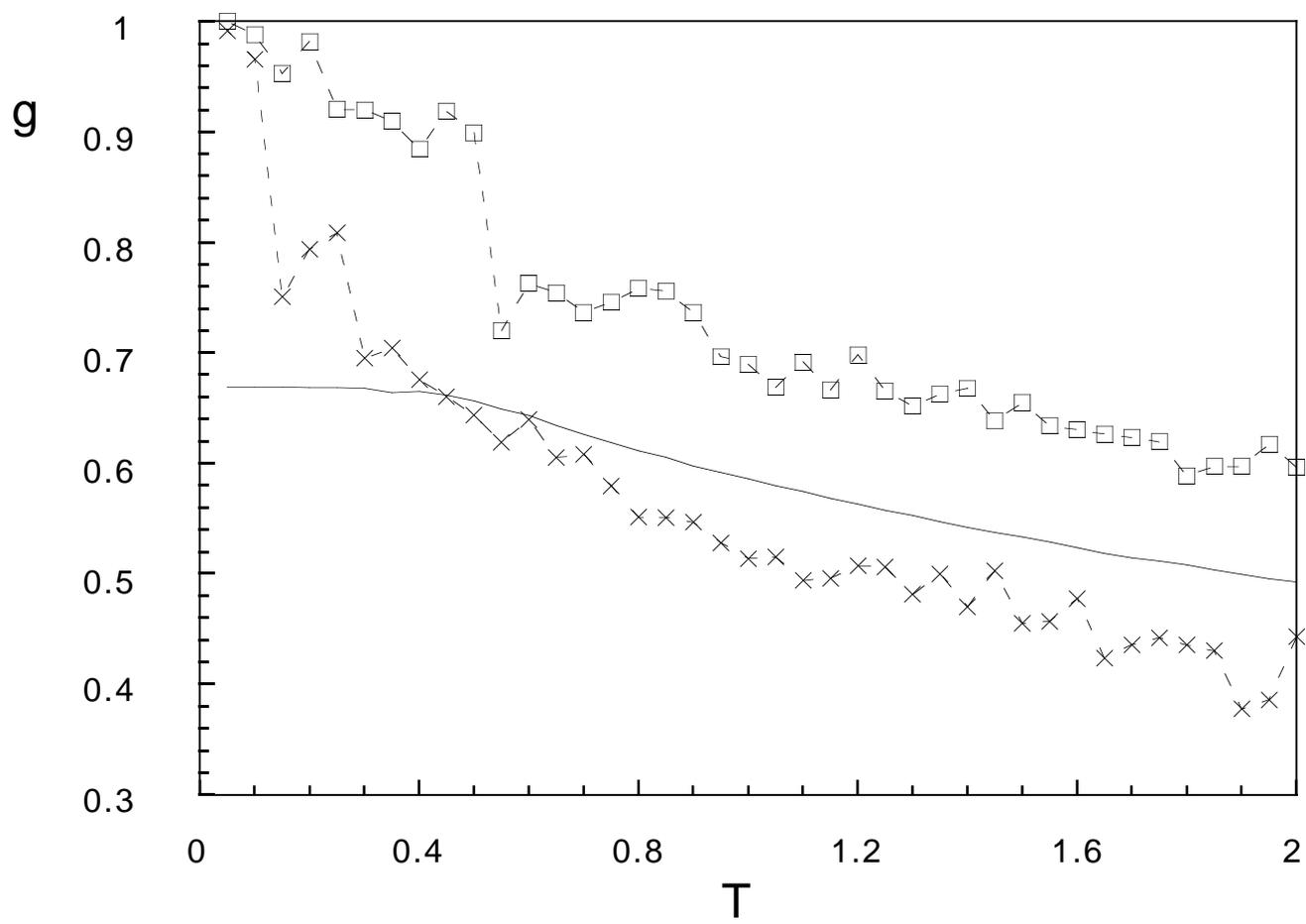
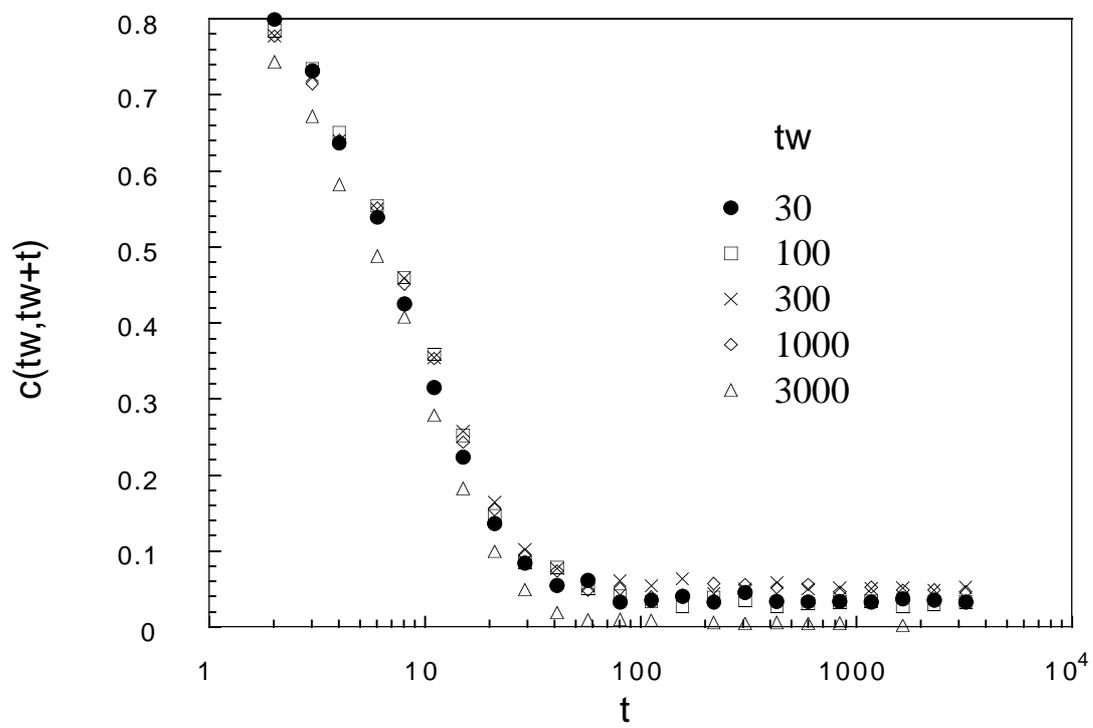
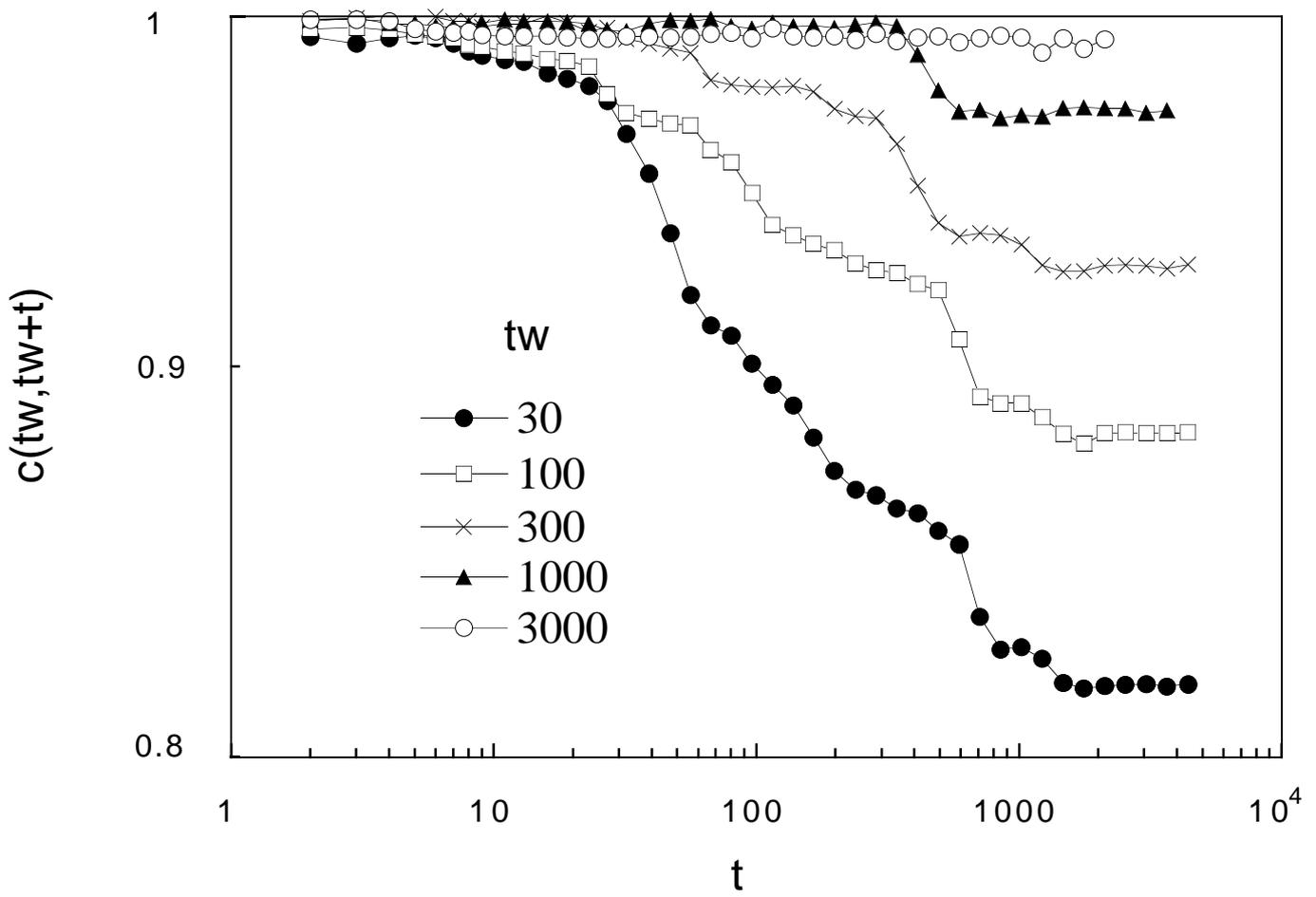
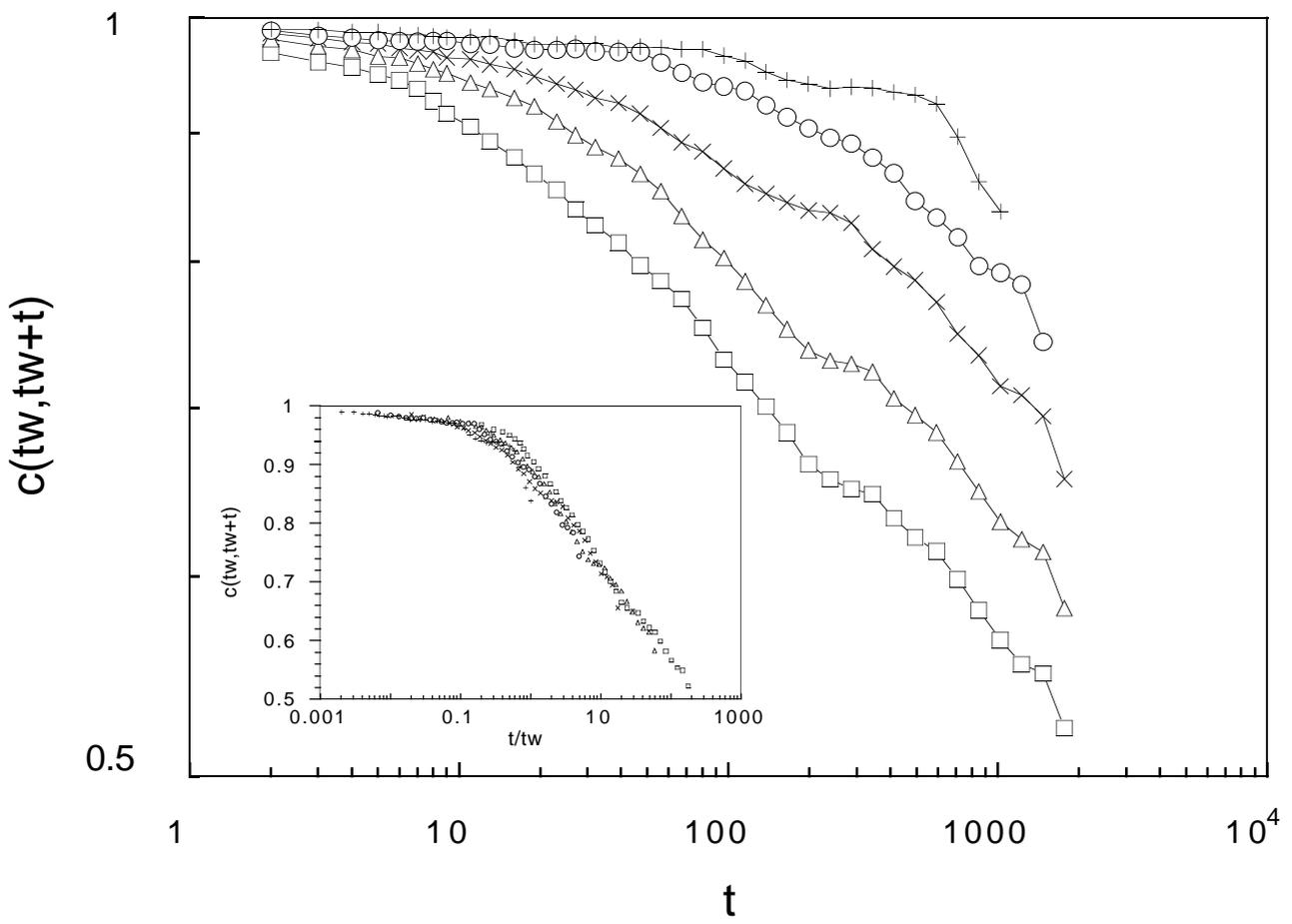
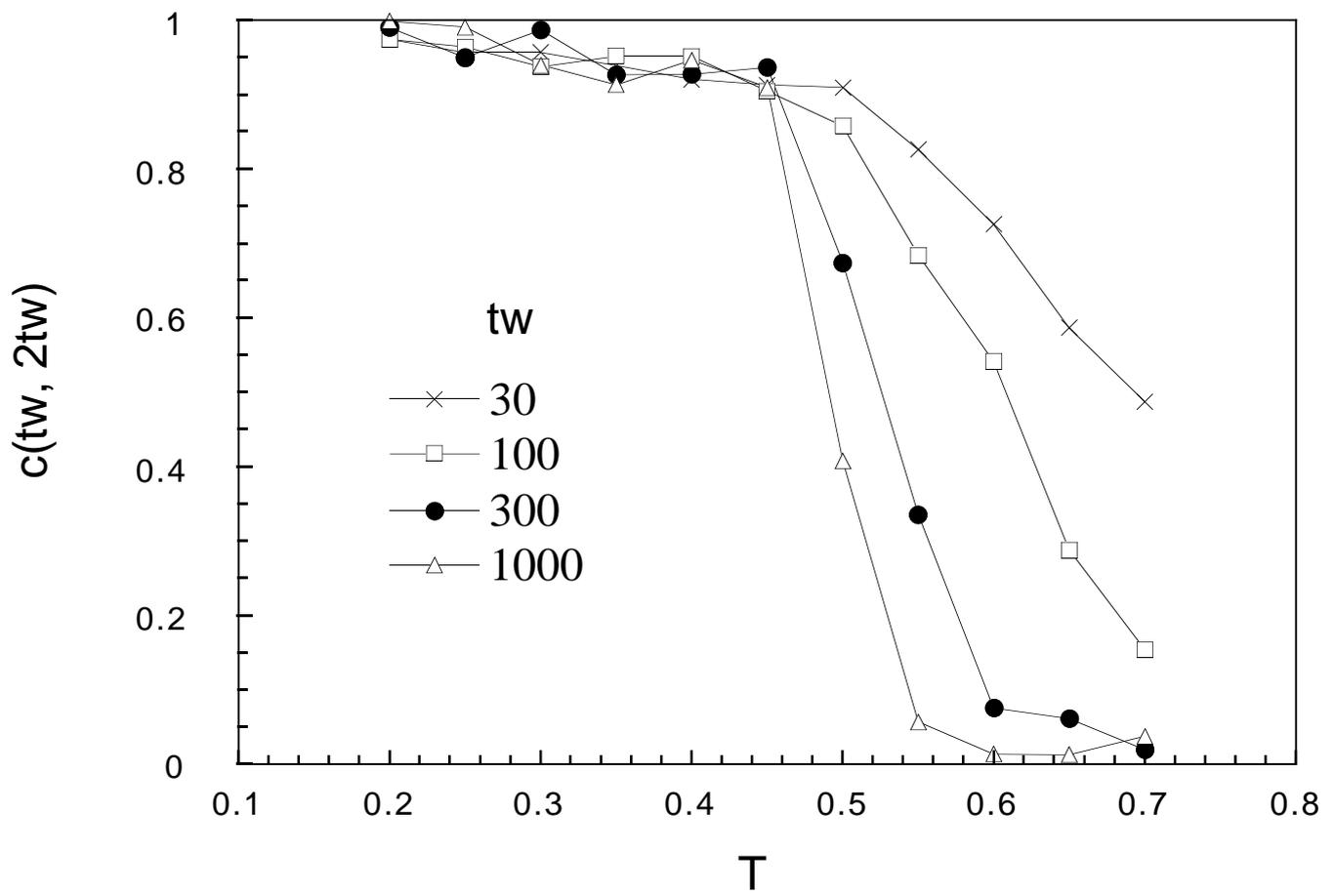
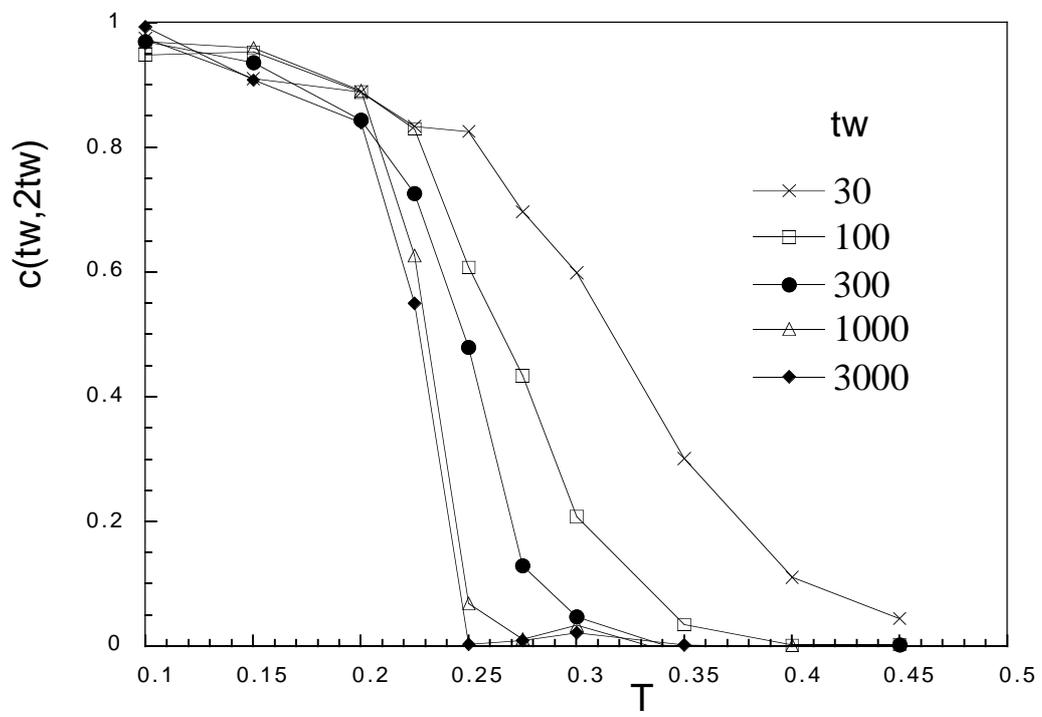
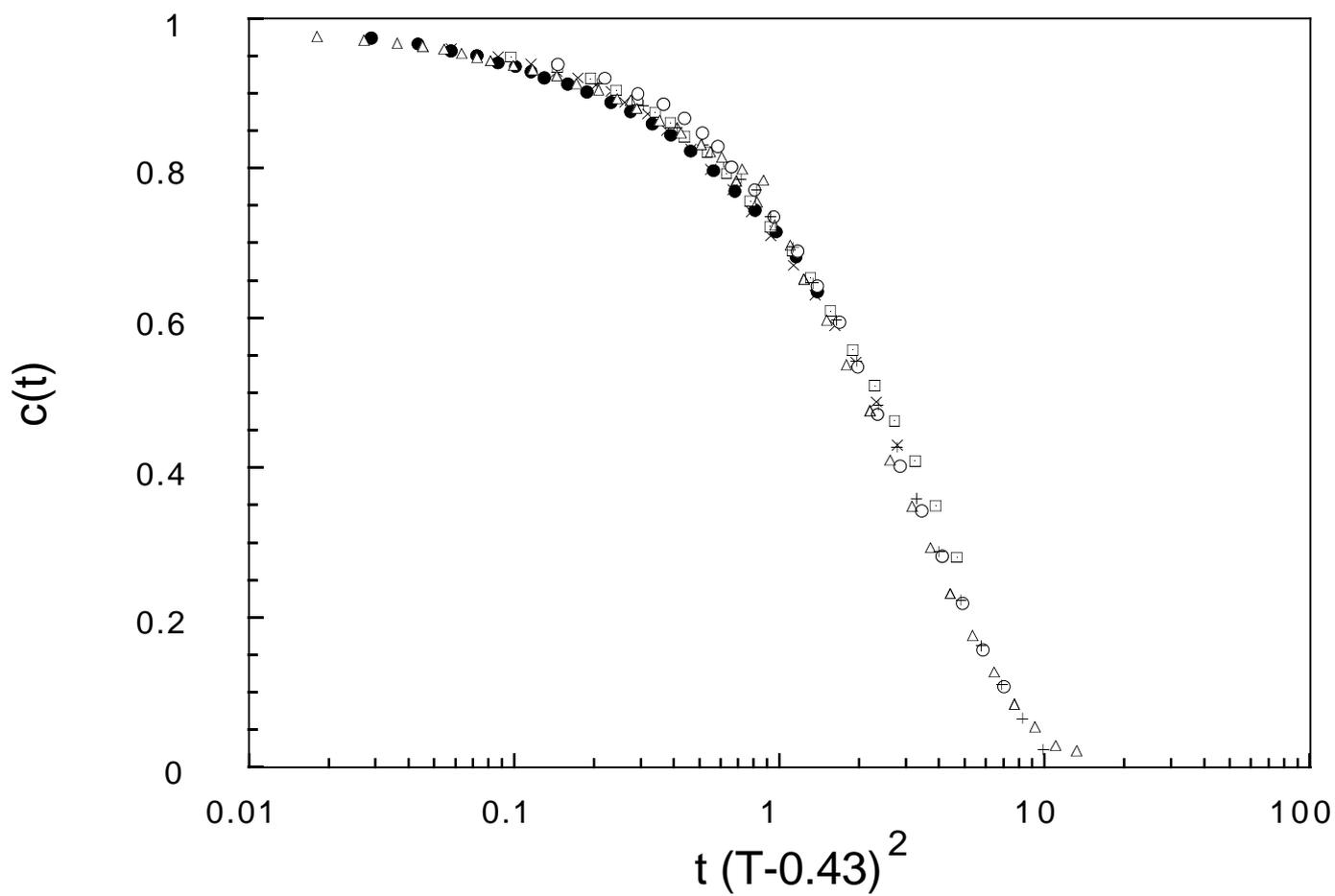
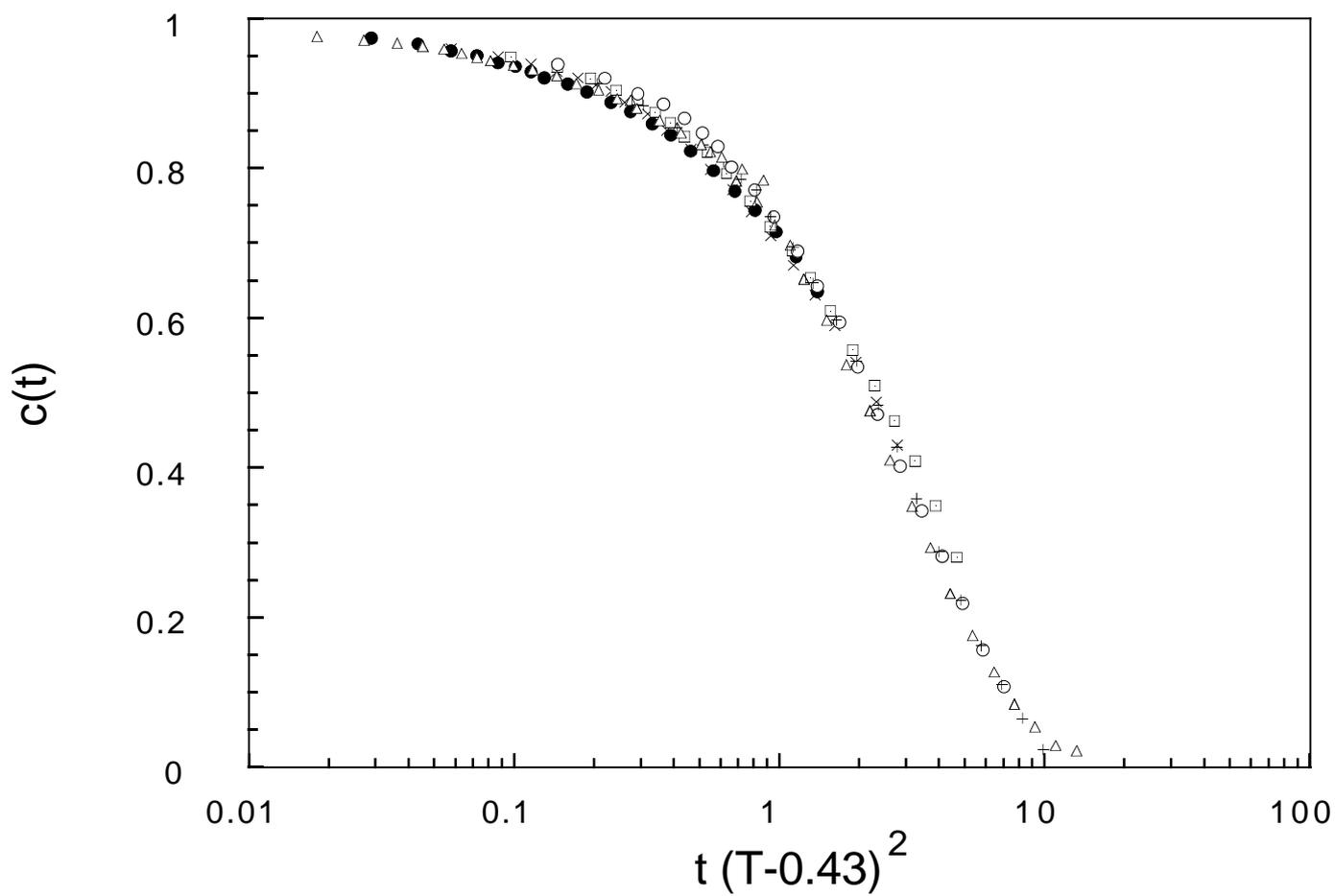
\end{document}